\documentclass[pdflatex,sn-mathphys-num]{sn-jnl}% Math and Physical Sciences Numbered Reference Style
\usepackage{graphicx}%
\usepackage{multirow}%
\usepackage{amsmath,amssymb,amsfonts}%
\usepackage{amsthm}%
\usepackage{mathrsfs}%
\usepackage[title]{appendix}%
\usepackage{xcolor}%
\usepackage{textcomp}%
\usepackage{manyfoot}%
\usepackage{booktabs}%
\usepackage{algorithm}%
\usepackage{algorithmicx}%
\usepackage{algpseudocode}%
\usepackage{listings}%
\usepackage{makecell}
\theoremstyle{thmstyleone}%
\theoremstyle{thmstyletwo}%

\theoremstyle{thmstylethree}%

\begin{document}

\title[The Schwurbelarchiv: A German-Language Dataset Derived from Multimodal Telegram Content for the Study of Conspiracy Theories]{The Schwurbelarchiv: A German-Language Dataset Derived from Multimodal Telegram Content for the Study of Conspiracy Theories}


\author[1]{\fnm{Mathias} \sur{Angermaier}}\email{mathias.angermaier@uni-graz.at}
\author[1]{\fnm{Elisabeth} \sur{Höldrich}}\email{elisabeth.hoeldrich@uni-graz.at}
\author[1]{\fnm{Joao P.} \sur{Neto}}\email{joaoxp@gmail.com}
\author*[1]{\fnm{Jana} \sur{Lasser}}\email{jana.lasser@uni-graz.at}

\affil*[1]{\orgdiv{IDea\_Lab}, \orgname{University of Graz},
\orgaddress{\street{Leechgasse 34}, \city{Graz}, \postcode{8010}, \country{Austria}}}


\abstract{
% !TEX root =  ../main.tex

Sociality borne by language, as is the predominant digital trace on text-based social media platforms, harbours the raw material for exploring a multitude of social phenomena. Distinctively, the messaging service Telegram provides functionalities that allow for socially interactive as well as one-to-many communication. The Telegram dataset presented here contains over 5,800 groups and channels discussing conspiracy-related topics with 63 million messages, originating from a data-hoarding initiative named the ``Schwurbelarchiv'' (from German schwurbeln: speaking nonsense). Uniquely, it includes the transcriptions of over 3 million audio and video files. 
Our contribution is a processed, research-ready version of this data hoard: we parse, clean, and validate the raw archive, pseudonymise user data, and transcribe roughly 126,000 hours of audio and video content. In its original form the archive was stored in a format that is difficult to process and largely inaccessible for systematic research.
This dataset publication details the structure, scope, and methodological specifics of the Schwurbelarchiv, emphasising its relevance for further research on the German-language conspiracy-related discourse. We validate its predominantly German origin by linguistic and temporal markers and situate it within the context of similar datasets. We describe process and extent of the transcription of multimedia files. Thanks to this effort the dataset uniquely supports analysis of text from originally multimodal sources like voice messages and videos to investigate online social dynamics and content dissemination. Researchers can employ this resource to explore societal dynamics for example related to conspiracy theories, misinformation, political extremism, and social network structures. 
%The Schwurbelarchiv thus offers unprecedented opportunities for investigations into digital communication and its societal implications.
}




\maketitle
% !TEX root =  ../main.tex

\section{Introduction}

The proliferation of digital trace data from social media platforms has enriched the study of societal phenomena, enabling researchers to analyse communication patterns and the dissemination of information at large scales \cite{freelon_interpretation_2014}. Platforms like Telegram, with its blend of interactive groups and broadcast channels, serve as fertile ground for examining issues ranging from political discourse to the spread of conspiracy theories. The insights gained from such analyses reveal their influence on societal narratives.

The \textit{Schwurbelarchiv}\footnote{The term \textit{schwurbeln}, while denoting pseudoscientific and irrational content more generally, gained widespread usage in German-speaking countries from 2021 onward in the context of the COVID-19 pandemic.} dataset we make accessible contributes to a growing number of published Telegram datasets for research, e.g.~\citet{la_morgia_tgdataset_2025, baumgartner_pushshift_2020, golovin2026teragram}.
It aims to provide a resource for researchers to enable large-scale investigations of the societal and linguistic dynamics that are connected with conspiracy theories, misinformation, political extremism, and social network structures, which has become a growing research interest in recent years~\cite{urman_what_2022, curley_covid-19_2022, bovet_organization_2022, hoseini_globalization_2023, willaert_disinformation_2022, zehring_german_2023, otherpaper}.
With the transcription of audio files we offer a unique starting point to widen the scope of research into these research areas by including multimedia content. 

We follow a layered definition of what constitutes a conspiracy theory, defined by the inclusion and exclusion criteria of content contained in the dataset (see Methods for details).
The original curator of the archive selected Telegram chats that ``\textit{address conspiracy myths / conspiracy theories / conspiracy fairy tales in some form. But also channels and groups from the surrounding milieu, e.g. vaccine passport sales}''~\cite{schwurbelarchiv}.
We validate this high-level definition by applying a machine-learning based classifier developed by~\citet{pustet_detection_2024} (see Methods for details). 
Accordingly, the granular definition for conspiracy theories we adopt here follows~\citet{pustet_detection_2024}: 
The term conspiracy theory is often used synonymously with disinformation, misinformation, rumours, or fake news~\cite{mahl2023conspiracy}.
While these phenomena can overlap, conspiracy theories have distinct features captured by the classifier: They assert a strong belief in a secret group intending to control institutions or even the world through intentionally causing complex, often unsolved events~\cite{mahl2023conspiracy, sunstein2009conspiracy}. 
Conspiracy theories offer alternative interpretations by attributing events to hidden powerful figures. 
They typically involve actors such as corrupt elites pursuing malicious goals, such as population control, through strategies like microchip insertion via vaccinations~\cite{samory2018government}. 
This is also compatible with the definition of~\citet{uscinski2018}, who defines conspiracy theories as explanations of past, ongoing, or future events that attribute their primary cause to a small group of powerful actors operating in secret for their own benefit and against the common good.

For empirical studies on conspiracy theories in German-language in particular, digital trace data from Social Media Platforms or Messenger Services have substantially contributed to the research landscape but still face limitations that we aim to address with the dataset described here. ~\citet{pustet_detection_2024} study conspiracy content in approximately 4,000 German Telegram messages and identify the scarcity of large-scale German language data as a limitation for the robustness of the research and its applicability to the broader discourse. They also note that existing datasets are often keyword-filtered instead of snowball-sampled, which creates systematic biases towards a few particular conspiracy theories and narratives.
Next to limitations of scale and sampling, the exclusive reliance on textual data leaves a substantial portion of Telegram communication unstudied. In the \textit{Schwurbelarchiv} dataset, media files such as images, videos, and voice messages make up 24.4\% of all messages. ~\citet{hartwig2024navigating} explicitly identify the neglect of voice messages in misinformation research as a critical gap, pointing to the absence of accessible voice message databases despite the growing role of audio content on platforms like Telegram and WhatsApp. Similarly,~\citet{steffen2025more} takes a first step towards multimodal analysis of German conspiracy channels by applying topic modelling to text-image pairs, yet notes that audio content remains entirely outside the scope of current approaches, despite its prevalence on Telegram. The \textit{Schwurbelarchiv} addresses these limitations by providing a large-scale, snowball-sampled, German-language dataset of over 63 million messages from more than 5,800 Telegram chats, combined with the transcription of roughly 126,000 hours of audio and video content, which allows text analyses of multimodal sources that move beyond the text-only, small-scale, and English-centric resources currently available.

The raw data contained in the \textit{Schwurbelarchiv} was not collected by the authors but was assembled by an anonymous individual and deposited on the Internet Archive. In its original form it comprises nearly 24\,TB of compressed multimedia files in formats that are not readily accessible for computational analysis. 
The contribution of our work is to present a processed version of the archive: we parse and clean the raw data, pseudonymise user information, assess the dataset's coverage, transcribe over three million audio and video files, and provide considerations regarding ethics and data protection, resulting in a structured, text-based dataset that is directly usable for computational social science research.

\section{Related Work}

Telegram is an instant messaging platform where users can exchange messages by creating an account with their phone number. Users can communicate in direct chats (one-to-one communication), in groups of up to 200,000 users (many-to-many), or participate in broadcast channels (one-to-many). These broadcast channels allow an unlimited amount of Telegram users to follow a single Telegram user's (the administrator's) content, and therefore broadcast channels are well designed for information dissemination at a large scale. While one-to-one communication is private and generally not accessible to researchers, groups and channels are public and accessible if their URL is known. We refer to groups and channels together as ``chats'' in the remainder of this work. Notably, Telegram does not offer a global search function for discovering chats by topic, which necessitates snowball sampling approaches as employed by the individual that curated the dataset described here.

A number of studies on Telegram datasets or research primarily using Telegram data have been conducted in recent years, many of which focus on the dissemination of misinformation and conspiracy theories in online environments. Here, we introduce the three Telegram datasets most closely related to our work in terms of temporal and linguistic scope of the data contained. We analyse similarities and differences across datasets, aiming to provide an assessment of the coverage of the \textit{Schwurbelarchiv} and situating the dataset within the relevant research context.

\subsection{Public Telegram Datasets.} The study by ~\citet{zehring_german_2023} investigates the connections between the Querdenken movement, a prominent anti-COVID protest network in German-speaking countries often linked to conspiracy theories, to far-right and alternative media, focusing on content-sharing patterns and the influence of QAnon within the network. Using a dataset of 6,294,955 text messages from 578 German public Telegram channels, the research explores the movement's key topics over time and across its sub-communities. The dataset was collected by first identifying seed channels by searching for actors connected to the Querdenken and far-right movements, to the esoteric/spirituality scene as well as nominees for the Golden Tin Foil Hat prize\footnote{\url{https://www.dergoldenealuhut.de}} in Germany, Austria, and Switzerland. This approach resulted in 238 initial channels, from which all messages were scraped. Next, frequently forwarded channels (mentioned 50 times or more) were identified and included in the dataset. This procedure -- also called ``snowball sampling'' -- was performed twice and led to a total of 578 public Telegram channels from which all messages available for the time between October 28, 2015 and January 3, 2022 were collected. By performing topic modelling on the dataset,~\citet{zehring_german_2023} found four overarching topics present in the Telegram channels: promotion, QAnon, right-wing populism, and COVID-19 conspiracy theories.

A larger dataset on similar topics was assembled by~\citet{golovin2026teragram}. Since snowball sampling on its own quickly yields an unmanageable amount of chats, they implemented an automated guidance mechanism that prioritises chats likely to contain relevant data based on a keyword list translated into 44 languages (the base list of English keywords is reproduced in Appendix~\ref{app:keywords}). The keywords targeted content containing misinformation related to COVID-19. The dataset, which is to our knowledge of the largest Telegram datasets available, includes 710,000 chats, comprising 5.9 billion messages, primarily in Russian, English, and German. The dataset contains text, metadata, and links, but excludes videos, images, and voice messages, which is a limitation that is central to the distinction with the \textit{Schwurbelarchiv}. The messages predominantly reflect activity from European time zones in the period from January 2015 to 2025, with trends linked to events such as the European COVID-19 vaccination campaign and the start of the Russian invasion of Ukraine. The dataset is available for researchers upon reasonable request. All predominantly German-language chats (23,000) from this dataset were made available to the authors of this study for comparison to the \textit{Schwurbelarchiv}.

The dataset collected by~\citet{baumgartner_pushshift_2020}, the largest publicly available collection of Telegram data to date, includes 27,000 chats and 317 million messages. The dataset is available for the years 2015-2019 and was regularly updated during that time, with source code available to facilitate further data collection, enabling researchers to extract information from specific chats. Starting with a seed list of 250 English-language chats -- 124 focused on right-wing extremism and 137 on cryptocurrency -- the number of chats included was also expanded using a snowball sampling approach. As for the dataset by~\citet{golovin2026teragram}, this dataset includes textual data only. Table~\ref{tab:comparison} provides a summary comparison of these datasets.

\begin{table}[ht]
\begin{center}
\renewcommand{\arraystretch}{1.4}
\resizebox{\columnwidth}{!}{
\begin{tabular}{|l|c|c|c|c|c|}
\hline
 & \textbf{Schwurbelarchiv} & \textbf{Zehring \& Domahidi} \cite{zehring_german_2023} & \textbf{Golovin} \cite{golovin2026teragram} & \textbf{Pushshift} \cite{baumgartner_pushshift_2020} & \textbf{TGDataset} \cite{la_morgia_tgdataset_2025} \\ \hline
Chats & 5,807 & 578 & 712,000 & 27,801 & 120,979$^*$ \\ \hline
Messages & 63M & 6.3M & 5.9B & 317M & 498M \\ \hline
Time span & 2015--2022 & 2015--2022 & 2015--2025 & 2015--2019 & 2015--2022 \\ \hline
Language & German & German & Multilingual & Multilingual & Multilingual \\ \hline
Topic focus & Conspiracy & Conspiracy / far-right & General & General & General \\ \hline
Transcribed voice messages & 65,433\,h & No & No & No & No \\ \hline
Transcribed videos & 61,406\,h & No & No & No & No \\ \hline
Publicly available & \makecell{metadata only,\\content upon request} & No & \makecell{metadata only,\\content upon request}  & Yes & Yes \\ \hline
\end{tabular}
}
\caption{Comparison of Telegram datasets. $^*$Channels only; no groups.}
\label{tab:comparison}
\end{center}
\end{table}

\subsection{Research on German-Language Conspiracy Content on Telegram.} Beyond the collection of datasets, a number of studies have used German-language Telegram data to investigate conspiracy-related content using digital-trace data and computational methods. ~\citet{weigand2022conspiracy} performed a long-term content analysis of conspiracy narratives in the German Querdenken protest movement, fine-tuning a DistilBERT model on self-annotated data from 34 chat groups comprising 822,000 text messages. ~\citet{pustet_detection_2024} compared supervised fine-tuning of BERT-based models with prompt-based approaches using GPT-4 and Llama~2 for detecting conspiracy theories in German Telegram messages, building on an annotated corpus of approximately 4,000 messages. More recently, ~\citet{steffen2025more} applied topic modelling on text-image pairs from 571 German conspiracy channels but still ignores voice messages and videos. These studies share a common set of limitations that the \textit{Schwurbelarchiv} dataset helps to overcome: They operate on comparatively small corpora and are limited to text or at most text and images. Audio and video content constitute a considerable share of Telegram messages~\cite{steffen2025more} (see also Table~\ref{tab:explo}), but has not been incorporated into any of these analyses. The \textit{Schwurbelarchiv}, with its scale and the transcription of audio files, is designed to overcome several of these limitations. For instance, the dataset enables investigating how conspiracy narratives evolve across thousands of groups over time, examining the role of voice messages and videos in disseminating conspiracy content compared to text, or tracing information flows through forwarding networks to understand how specific narratives spread between groups as demonstrated by~\citet{otherpaper}.

\section{Methodology}
\subsection{Data Collection.}
The entire raw data contained in the \textit{Schwurbelarchiv} is available on the Internet Archive\footnote{\url{https://archive.org/details/schwurbel-archiv}}\footnote{\url{https://web.archive.org/web/20241126233137/https://schwurbelarchiv.wordpress.com/}}. The person initially responsible for collecting the data remains anonymous, but responded to requests by the authors via email and provided information about the data collection process. They describe their motivation for collecting the dataset as the pure joy of data-hoarding, contribution to the Internet Archive, and enablement of research and journalism using the collected data as a source. 

Similar to the other Telegram datasets described here, groups and channels in the \textit{Schwurbelarchiv} were selected via a snowball sampling approach. Starting with a seed set of Telegram chats focused on renowned German-speaking conspiracy theorists, other chats were identified via invitation links and forwarded messages originating from the chats in the sample. Newly discovered chats were added to the sample, and all messages posted in them up to the point of their discovery were scraped via bulk download, after which a continual streaming process captured new content in near real-time. Relevance of newly discovered chats was determined by manual curation (e.g., human in the loop) that excluded chats deemed unrelated to conspiracy theories in order to prune irrelevant sampling paths early. Data collection was terminated when the anonymous collector lacked the resources to continue. The resulting dataset covers the time span from September 23, 2015 to August 5, 2022. 

Snowball sampling promises to capture a more representative portion of the discourse but still has a number of limitations: manual curation of discovered chats ensures relevance but remains subjective, relevant chats might not be included because sampling terminated before a link to them was found, and relevant chats might be missed because they are entirely disconnected from the forwarding network in which sampling was initiated. In particular, chats from which few messages are forwarded are systematically less likely to be discovered by this approach. We note, however, that snowball sampling is the only available methodology for data collection from Telegram beyond manual search and selection, as the platform lacks a searchable index of public chats. 

In contrast to the data collection approach of the other studies described here which relied on the Telegram API, the data in the \textit{Schwurbelarchiv} was gathered using a Windows Virtual Machine, with four separate Telegram Desktop sessions, each operating under its own account. Chats were then exported using the Remote Desktop Protocol. Once a chat has been discovered by the snowball sampling process, all its content up to that point in time was bulk-downloaded. Notably, this approach also enables streaming of the content after a chat has been discovered. This is important because the deletion of messages by users on Telegram has emerged as a significant factor influencing the completeness and reliability of Telegram datasets. \citet{buehling_message_2024} highlight how this can lead to systematic biases in analyses such as topic modelling, dictionary-based computational content analysis, and network studies. The study finds that for channels, on average, only 88\% of messages remain accessible five days after posting, dropping to 83\% after seven months. In groups, the decline is even steeper, with only 64\% of messages remaining after five days and 52\% after seven months, emphasizing the need for timely data collection. 
%While we do not know at which point a given chat was discovered and its content streamed, the \textit{Schwurbelarchiv} is potentially more complete when compared to other datasets regarding message deletions.

The Remote Desktop Protocol enables storing not only the HTML text files, but all video files, voice messages, and images posted in the chats. Video files larger than 8\,MB have not been downloaded until mid of 2021, whereafter the limit for downloaded videos was increased to 50\,MB. To our knowledge, the \textit{Schwurbelarchiv} is the only  Telegram dataset available for research that contains multimedia content. This content constitutes a large part of the discourse taking place on social media platforms~\cite{steffen2025more} that is typically ignored due to the difficulty of handling diverse data formats and large data volumes. The fact that the dataset we provide retains this content and makes the multimodal files accessible through transcribing them into text-data greatly expands the types of research questions and analysis approaches that it enables. 

All data columns describing message content and metadata (see Table~\ref{tab:columns}) were parsed from the HTML files exported by the original data collector. The columns \texttt{message\_lang}, \texttt{audio\_lang}, and \texttt{video\_lang} were added by us using the Lingua language detection package~\footnote{https://github.com/pemistahl/lingua-py}, \texttt{is\_conspiracy} was computed using the fine-tuned BERT model by \citet{pustet_detection_2024} as described in Section ``Conspiracy Content in Telegram Chats'' below, and all transcription-related columns (\texttt{audio\_transcription}, \texttt{video\_transcription}, and associated duration and file size fields) were generated by us using the Whisper model described in the following section.

\subsection{Data Cleaning and Preprocessing}
Since the raw collected data contains multimedia files, it requires a storage space of nearly 24 TB in a compressed state. The sheer size of the data as well as the native Internet Archive file format \texttt{.warc} complicate processing and analysis. Therefore, next to clarifying ethical, data protection-related, and licensing questions associated with the use of the \textit{Schwurbelarchiv} for research (see Section ``Ethics and Data Protection'' below), we set out to preprocess the dataset to make it more easily accessible for a wider research community. This preprocessing involved data cleaning, transcription of audio content, and an assessment of the completeness of the dataset. We describe these steps in more detail in the remainder of this section. The fully documented pipeline is available on GitHub\footnote{https://github.com/cs2-lab/schwurbelarchiv}.

After downloading the raw data from the Internet Archive, we used Python scripts to extract all relevant information, namely \texttt{author}, \texttt{posting date}, and \texttt{message body}. In addition, we distinguish between original messages and forwarded messages -- and their respective authors and posting dates -- which enables the reconstruction of message forwarding networks within the dataset. Furthermore, we isolate links to websites shared in the message text body. A full listing of all data columns contained in the cleaned dataset is provided in the Appendix~\ref{tab:columns}. While the raw dataset as it is published on the Internet Archive contains uncensored usernames, we pseudonymize usernames using a SHA-256 hash-function. We extract usernames from the HTML files containing the content of the chats but do not have access to user IDs assigned by Telegram. As a result, we cannot distinguish between users that have chosen the same username. This is a substantial limitation of the dataset at hand. We further note that Telegram allows the use of bots as automated accounts, which may also be present in the dataset. We did not attempt to identify or flag bot accounts, which constitutes another limitation of the dataset.

In addition to textual data, we transcribed the audio of both audio and video files. We used ffmpeg to first transcode the various audio codecs into WAV, and \texttt{openai/whisper-large-v3-turbo}\footnote{https://huggingface.co/openai/whisper-large-v3-turbo} to transcribe the audio content. Various optimizations were used in order to make the transcription task more computationally efficient, including using the CTranslate2 faster-whisper implementation\footnote{https://github.com/SYSTRAN/faster-whisper} with Voice Activity Detection and batching. These resulted in a transcription real time factor of $\sim  250$X per Nvidia L40S GPU. Overall, 126,839 hours of content (14.9 years) were transcribed.

During the initial data collection, there was no quality control of the included chats with respect to research interests such as the investigation of temporal conversation patterns. However, the raw dataset contains a substantial number of ``dumps'' -- chats that were started anew by dumping all the content of a discontinued chat into them. Such chats add redundant information to the dataset, might artificially skew the relation of content posted per author, and distort analyses that leverage the time stamps of messages to reconstruct temporal patterns. To remove such dumps, we therefore excluded chats that either showed activity for a period shorter than 10 days or where more than 90\% of the content included was posted in a single day. In total, we excluded 1,050 such dump chats (13.7\% of total) with 317,532 messages (0.5\% of total). 
While we do not further use the content from these chats, their sheer existence provides an estimation of how many chats in the domain of conspiracy discussions on Telegram are discontinued. After cleaning the dataset, it contains a total of 63,908,365 messages posted in 5807 unique chats, covering the time span from 23/09/2015 to 05/08/2022 (see also Tab.~\ref{tab:explo}).

\subsection{Ethics and Data Protection}
While the data contained in the \textit{Schwurbelarchiv} was collected from publicly accessible chats, it should not be considered ``fair game'' for research without consideration for the privacy expectations of the users whose messages are contained in the dataset. In preparing the dataset for publication, we considered a number of arguments to weigh the potential infringement on user expectations against the merit of potential research enabled by making the data accessible, which we will detail in this section. 

Telegram users can freely choose their username on the platform. Therefore, more than one user can choose the same username and usernames do not necessarily allow the identification of individuals outside Telegram. Nevertheless, many usernames contained in the meta information of posts resemble names of real people. Even though the content contained in the dataset is public, users of social media platforms usually do not expect their posts to be singled out and discussed in research projects~\cite{townsend_social_2016}. We therefore argue that any attempt at linking Telegram posts to the identities of people outside the platform would be problematic from the perspective of user expectations. Therefore, we pseudonymized all messages by replacing usernames with unique IDs via hashing. As a result, messages that come from the same username can still be linked, but any resemblance of usernames to real names is removed. We believe that this step does not hinder the exploration of discussion topics and communication dynamics while it helps to balance user expectations with the potential merit of using the \textit{Schwurbelarchiv} for research.

A similar consideration has to be made for the names of groups and channels in the dataset. While the majority of these names refer to places or discussion topics, a number of chat names resemble the names of real people. This is particularly frequent for channels where single individuals post content that is broadcast to the followers of the channel. For the present data publication, we chose to retain the names of chats and not pseudonymize them. This decision is driven by two main considerations: First, by altering chat names, analyses like the identification of the geographical scope of chats that relies on chat names would not be possible. Second, users that choose to create a chat under their name expect to receive a substantial amount of attention. Following ~\citet{townsend_social_2016}, we argue that it is ethically admissible to retain these names, as providing them in the dataset will neither significantly heighten the attention these users receive nor breach their expectations.

Even though usernames are anonymized, the content of the posts might allow for the re-identification of people based on disclosed attributes such as location, demographic characteristics, or real names that are contained in message texts rather than usernames. Pseudonymization of this information in message texts let alone multimedia content is not technically feasible. However, re-identification of individuals at scale would require some effort to parse content and link user identities to messages. While we consider this technically feasible for individual messages, we caution against any such attempts to honour expectations of individual users. To limit the risk of re-identification, the dataset we openly publish on Zenodo contains only metadata and excludes all free-text content, namely the text of messages and forwarded messages together with their speech transcriptions~\cite{data_schwurbel_clean}. The full dataset, which additionally includes these texts~\cite{data_schwurbel_full}, is not distributed openly but published as a separate, access-restricted record: interested researchers can request access directly through Zenodo, stating their identity and intended use. Access will be granted by the authors on a case-by-case basis.

While the anonymous collector of the \textit{Schwurbelarchiv} explicitly states their intention for the usage of the dataset as a source for journalists and research on the website of the dataset\footnote{https://web.archive.org/web/20241126233137/}, neither the website nor the Internet Archive collection contain a license. This in principle precludes re-use and re-publication of the dataset. However, in private conversation with the authors of the present work, the anonymous collector confirms that they intended to publish their data under a CC0 license and explicitly permit re-use and re-publication. 

Lastly, we obtained IRB clearance for research with and re-publication of the data contained in the \textit{Schwurbelarchiv} (vote of the Institutional Review Board of Graz University of Technology from March 28, 2023). 

\section{Dataset Description}

In the following section, we first describe the \textit{Schwurbelarchiv} in more detail, giving statistics on the different content types it contains as well as linguistic and temporal markers. %In addition, we provide an analysis of overall posting activity and activity distribution across authors and chats. 
After establishing basic characteristics of the dataset, we describe our efforts to assess the completeness of the dataset and the extent to which it relates to conspiracy theories. Lastly, we provide more detail on the results of the audio content transcription.

\subsection{Descriptive Statistics}
Table~\ref{tab:explo} shows descriptive statistics of the cleaned \textit{Schwurbelarchiv} dataset. With the ratio of channels to groups in this dataset being roughly 2:1\footnote{We use HTML-parsing heuristics to determine the category of a chat, which resulted in a small number of chats (37) to be manually determined.}, the \textit{Schwurbelarchiv} dataset can for example be used to investigate whether conspiracy-theory-related discussion arises organically through conversation in discussion groups, or whether agenda-setting by a small number of users in broadcast channels is more prevalent. We find that media files like \textit{images}, \textit{videos}, and \textit{voice messages} make up 24.4\% of all messages. While they account for 42.2\% of all forwarded messages, for original messages only 16.7\% are media files. This difference makes the analysis of the content of media files particularly relevant for the investigation of content dissemination. The distinction between forwarded messages and original messages also serves to estimate the scope of the conspiracy related German-language Telegram discourse, which we will leverage in Section ``Validation and Completeness'' below to gauge the completeness of the dataset. 

\begin{table}[!ht]
\begin{tabular}{|c|c|c|c|} \hline
                   & Original   & Forwarded  & Total      \\ \hline
Total chats             & 5,807       &            &            \\ \hline
Channels           & 3,486       &            &            \\ \hline
Groups             & 2,321       &            &            \\ \hline
Authors            &  348,812&  611,164& 896,339\\ \hline
Content            & 43,739,502& 18,855,979& 62,595,481\\ \hline
Text               & 38,021,236& 15,797,326& 53,818,562\\ \hline
Images             & 5,707,321& 6,166,370 & 11,873,691 \\ \hline
Video              & 1,065,146  & 1,431,740 & 2,496,886  \\ \hline
Voice messages     & 542,928    & 354,531 & 897,459    \\ \hline
External links& 6,965,776& 5,356,078&12,321,854\\ \hline
\end{tabular}
\caption{Descriptive statistics of the content of the \textit{Schwurbelarchiv}. A small number of chats (15) were of an unknown category.}
\label{tab:explo}
\end{table}

\subsection{Language, time zone and region} 

\begin{figure}[ht]  
\centering
\includegraphics[width=\columnwidth]{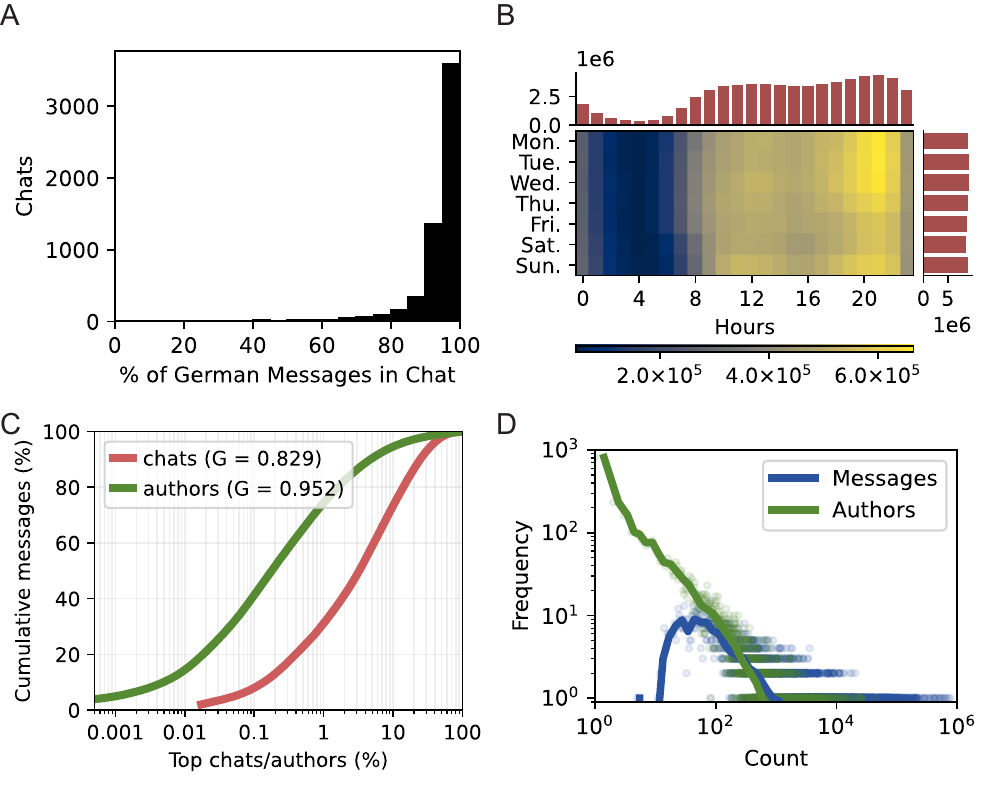}  
\caption{
\textbf{A.} Percentage of German messages in chats.
\textbf{B.} Posting frequency aggregated by hour of the day and day of the week in the UTC+1 time
zone.
\textbf{C.} Frequency distribution of messages and authors per chat. G indicates the Gini coefficient of the cumulative distribution.
\textbf{D.}  Lorenz plot showing cumulative fraction of messages per top chats and authors. 
}
\label{fig:dataset_statistics}
\end{figure}

To evaluate the language distribution within the chats in the dataset, we classify all messages with the Lingua package. Filtering out empty, short (less than 5 characters, e.g. emoji replies) and URL-only messages we are left with 52.5\% of all messages. From those, German is by far most prevalent, with 91.3\% of the messages followed by English with 2.2\%. We also examined the distribution of German message proportions across chats, shown in Figure~\ref{fig:dataset_statistics}A. The vast majority of chats (98.6\%) contain at least 50\% of messages in German, while in 85.5\% of chats it exceeds 90\%.

As messages on Telegram are not geo-tagged, we have no direct information about the geolocation of the authors that posted messages contained in the dataset. However, we can use other information to narrow the geographic scope of the observations. In Fig.~\ref{fig:dataset_statistics}B we show the posting pattern on different weekdays and hours of a day using the UTC+1 time zone (e.g., Central Europe). We observe a consistent minimum in posting frequency during the European night hours with maxima in posting frequency in the early evening hours and on Sundays. This observation indicates that the diurnal rhythm of the majority of authors in the dataset is consistent with Central Europe.

% \begin{table}[h]
% \centering
% \resizebox{\columnwidth}{!}{%
% \begin{tabular}{|c | c c|} 
%  \hline
%   & Number of Regional Related Chats & \% of full dataset\\ [0.5ex]  \hline
% Schwurbelarchiv & 1,514 & \textbf{24.9 \%} \\  \hline
% Mohr &2,275 &9.9  \%  \\ \hline
% Zehring & 33 & 5.7 \% \\ \hline
% \end{tabular}
% }
% \caption{Comparison of regional Telegram chats across datasets.}
% \label{tab:full_regionality}
% \end{table}

\subsection{Validation and Completeness of the Dataset}
In light of the absence of a structured database of Telegram content that can be used to retrieve a comprehensive dataset for a given query, it is important to assess the completeness of collected Telegram datasets and its relatedness to the targeted scope, e.g. German-language conspiracy-related discourse. Since Telegram does not provide a searchable index of public chats nor any official record of chats about specific topics, no ground truth exists with which the completeness of any Telegram dataset could be assessed. All existing Telegram datasets therefore rely on manual search and selection or snowball sampling. As discussed in Section ``Data Collection'' above, snowball sampling is systematically biased toward including chats with higher forwarding activity, meaning that smaller or more insular chats are likely underrepresented in the dataset. The metrics we report can therefore provide only an approximate assessment of completeness.
To this end, we consider alternative means of studying completeness, using the information that is available about the dataset and in the published literature: The anonymous collector of the dataset states that the \textit{Schwurbelarchiv} contains conspiracy-related content (see also Introduction). We start by validating this claim via classification of message content.
Next, and motivated by the reported frequent message deletions on Telegram~\cite{buehling_message_2024}, we gauge the probability of content deletions to have taken place before data was collected given the data collection approach employed. 
Next, we attempt to gauge the extent to which the dataset covers the discourse that is in scope of the sampling. While no searchable index of content exists, forwarded content provides a means of assessing the proportion of the targeted discourse that is included in the sample by analysing the presence of original messages and forwarded messages: if many forwarded messages are not present in the dataset as original messages, this is evidence that a substantial part of the discourse is not captured. 
Lastly, we employ a complementary approach to assess completeness of the dataset by looking at the chats that the \textit{Schwurbelarchiv} includes. Since there is no official record of all German-speaking Telegram chats dedicated to conspiracy-related content we compare the chats in the dataset with chats in similar datasets, specifically those  by~\citet{zehring_german_2023}, a German-language subset of \citet{golovin2026teragram},  and~\citet{baumgartner_pushshift_2020} introduced earlier.

\subsubsection{Conspiracy Content in Telegram Chats}
As described in the Methods Section above, the dataset was collected anonymously through snowball sampling of conspiracy-related Telegram groups with the primary aim of capturing conspiracy-related content on Telegram. However, other than manual curation by the data collector, no validation was conducted to ensure that the dataset effectively captured conspiracy content, nor was there an assessment of the proportion of messages within the chats that actually contain such content.  

To validate the scope of the dataset, we applied TelConGBERT, a fine-tuned German BERT model (\texttt{deepset/gbert-base} with additional in-domain pre-training) developed by~\citet{pustet_detection_2024}, which assigns a binary label indicating whether a message communicates conspiracy-related content. The model operationalises conspiracy theories as the assertion that a secret group of powerful actors deliberately causes complex, typically unresolved events in pursuit of a malicious goal, structured around a good--evil dualism and the narrative components of actor, strategy, and goal; notably, only messages expressing belief were treated as positive instances, while mere references (e.g.\ hashtags such as \#NWO) were excluded to prevent the classifier from keying on explicit signal terms. On its test set the model attains an $F_1$ of $0.79$ for the positive class and a macro-averaged $F_1$ of $0.85$. TelConGBERT was trained on the TelCovACT corpus of COVID-19-era German Telegram messages, the same content, language, time- and platform domain that the data in the \textit{Schwurbelarchiv} is drawn from. The results on the Schwurbelarchiv corpus showed that 18.4\% of chats (1055 chats) do not contain conspiracy-related content. However, these chats account for only 0.5\% of all messages longer than 150 characters. This indicates that the majority of the dataset indeed contains conspiracy-related material.

\subsubsection{Deleted messages}
Assessing the completeness of the dataset on a message level involves examining the potential for missing or deleted messages, which is closely tied to the data collection approach. There is no published metadata on the time of data collection. In particular, the start of the streaming process (after the initial bulk download upon discovery, see Methods for details) for each chat is not known. However, the difference between bulk download and streaming has direct consequences for the expected completeness of messages in the dataset at different points in time: In the bulk download phase, it is probable that the dataset is missing messages that were deleted by authors prior to the download and that the percentage of missing messages increases further back in time from the time of the download. These deleted messages have no placeholders in the dataset and are therefore not visible. In contrast, during the continual streaming phase, messages are captured in near real-time, therefore reducing the probability of missing messages caused by user deletions. We were not able to detect the transition point where the data collection method changed from bulk to stream download for each chat. Therefore, we cannot guarantee the completeness of the dataset on a message level with respect to message deletions, since we cannot indicate the presence or fraction of deleted messages. However, it is probable that message deletions are less likely to have occurred for the later years contained in the dataset and that the \textit{Schwurbelarchiv} is less affected by message deletions than other comparable datasets.

Even though the dataset likely has less deleted messages, researchers using the data should still consider the systematic bias and possibly skewed results in computational analysis introduced by message deletions, as stated by \citet{buehling_message_2024}. These findings indicate a key limitation of the dataset provided here, since the lack of metadata detailing the data collection process prevents us from determining the extent to which deleted messages may have influenced its completeness.

\subsubsection{Forwarded Messages}
To further assess the completeness of conspiracy-related chats in the dataset, we performed a comparative analysis of forwarded and original messages. In the context of Telegram, a \textit{forwarded message} refers to a message which has been forwarded from another chat and is marked as such in the chat it is forwarded to. These forwarded messages can serve as proxies for content influence, reflecting the flow of information between chats. We calculated the proportion of forwarded messages which are also present in the dataset as original messages. This metric indicates the degree to which the original sources of forwarded messages are represented in the dataset. If this value were 100\%, it would imply that all forwarded messages are also present in the original chat the message was posted in, achieving complete coverage of the source population that is in scope of the sampling. In this dataset, this proportion was calculated as the fraction of forwarded messages whose original sender (fwd\_author) and original posting time (fwd\_posting\_date in UTC) jointly match an original (non-forwarded) message in the dataset (matching on author and posting\_date after UTC normalisation). In the dataset this value is 54.1\%. This finding suggests that we cover about half of the German-language conspiracy related discourse on Telegram in the observation time frame, highlighting the dataset's limitations in capturing the full extent of the discourse. We interpret this proportion as a rough estimate of coverage. A higher or lower value is not inherently desirable, as it reflects the scope of the sampled forwarding network rather than the quality of the data contained or the representativeness of the sampling.

\subsubsection{Chats Covered}
Lastly, we also compare the contents of the \textit{Schwurbelarchiv} to the chats collected by~\citet{zehring_german_2023}, a German-language subset of \citet{golovin2026teragram}, and~\citet{baumgartner_pushshift_2020}. 
The dataset by~\citet{zehring_german_2023}, consists of 578 chats and covers the time period October 28, 2015, to January 3, 2022. Among these chats, there are two that we have excluded from this dataset due to their classification as \textit{dump} chats. Out of the remaining chats, 121 are also present in the \textit{Schwurbelarchiv}, representing approximately 22\% of Zehring's chats and 2\% of the chats contained in the \textit{Schwurbelarchiv}. Although this overlap is smaller than expected given the similar selection criteria and seed topics, several factors may account for the discrepancy: \citet{zehring_german_2023} concentrated their seed selection on groups with substantial societal influence, including opinion leaders, politicians, and well-known conspiracy theorists. In contrast, the \textit{Schwurbelarchiv} dataset includes numerous smaller, more specialized groups often focused on local issues. For example, groups such as \textit{Corona Rebellen} have multiple sub-groups tailored to different cities and regions, reflecting a tendency towards decentralization in the \textit{Schwurbelarchiv} dataset. 
%This is also reflected in the overall share of content we identified as coming from regional groups, which is much higher than for~\citet{zehring_german_2023} (see also Tab.~\ref{tab:full_regionality}).

The subset of German-language chats in the dataset collected by~\citet{golovin2026teragram} consists of 22,968 chats. Among these, 46 have been categorised as \textit{dump} chats by the authors and have been excluded. From the remaining chats, 801 are also contained in the \textit{Schwurbelarchiv}. This corresponds to 13\% of the chats in the \textit{Schwurbelarchiv}, while only comprising 3\% of chats contained in the dataset collected by~\citet{golovin2026teragram}. 

The Pushshift dataset~\cite{baumgartner_pushshift_2020} is the most comprehensive publicly available dataset to date, encompassing over 27,000 chats from various countries. Of these, 78 chats (1.3\%) overlap with the \textit{Schwurbelarchiv}. None of the excluded dump chats in the \textit{Schwurbelarchiv} are present in the Pushshift dataset either. This low overlap is likely due to linguistic and regional differences in the selection criteria: while the \textit{Schwurbelarchiv} and datasets by~\citet{zehring_german_2023} and~\citet{golovin2026teragram} focus on German-speaking chats, the Pushshift dataset has a more global scope. In addition, the temporal scope of data collection between the Pushshift dataset (2015-2019) and the \textit{Schwurbelarchiv} (2015-2022) differs substantially.

%\begin{table}[ht]
%\begin{center}
%\resizebox{\columnwidth}{!}{
%\begin{tabular}{|l l | c c c|}  \hline 
%Dataset & & Number of Chats & Overlap with \textit{Schwurbelarchiv} & Overlap in \% \\  \hline
%Mohr & \cite{mohr_inference_2023} &22,968 &801 &13.0 \%  \\ \hline
%Zehring \& Domahidi & \cite{zehring_german_2023} & 578 & 121 & 2.0 \% \\ \hline
%Pushshift & \cite{baumgartner_pushshift_2020} & 27,801 & 78 & 1.3 \% \\ \hline
%\end{tabular}
%}
%\caption{Number of chats per dataset and overlap with the 5,963 chats in the \textit{Schwurbelarchiv}.}
%\label{tab:overlap}
%\end{center}
%\end{table}

Table~\ref{tab:overlap} summarizes the number of chats in each dataset and their overlap with the \textit{Schwurbelarchiv}, reported both in absolute numbers and as a percentage of \textit{Schwurbelarchiv} chats. Furthermore, the overlap between the~\citet{zehring_german_2023} and German-language subset of the~\citet{golovin2026teragram} datasets is 369 chats, corresponding to 66\% and 1.6\% of the content of each of these datasets, respectively. A total of 102 chats are present in all three datasets. Additionally, 39 chats are shared between the subset of the~\citet{golovin2026teragram} and Pushshift datasets, while no chats overlap between the datasets by~\citet{zehring_german_2023} and Pushshift.

\begin{table}[ht!]
\begin{center}
\renewcommand{\arraystretch}{1.4}
\resizebox{\columnwidth}{!}{
\begin{tabular}{|l l | p{5.5cm} | c c c|}  \hline 
Dataset & & Description & Number of Chats & Overlap & Overlap in \% \\  \hline
German-language subset of Golovin et al. & \cite{golovin2026teragram} & COVID-19 misinformation; multilingual; keyword-guided snowball sampling & 22,968 & 801 & 13.0\% \\ \hline
Zehring \& Domahidi & \cite{zehring_german_2023} & German Querdenken movement; far-right and conspiracy content & 578 & 121 & 2.0\% \\ \hline
Pushshift & \cite{baumgartner_pushshift_2020} & General; multilingual; seeded from right-wing extremism and cryptocurrency channels & 27,801 & 78 & 1.3\% \\ \hline
\end{tabular}
}
\caption{Number of chats per dataset and overlap with the 5,963 chats in the \textit{Schwurbelarchiv}.}
\label{tab:overlap}
\end{center}
\end{table}

Lastly, we compared the \textit{Schwurbelarchiv} dataset with the channels and groups monitored by the \textit{Machine Against the Rage} initiative by \textit{BAG Gegen Hass im Netz} \cite{machinerage2024}, which shows network patterns of forwarded messages within the ideological spectrum of conspiracy theories and extremism. We found an overlap of 212 chats with the Schwurbelarchiv. The initiative's monitoring begins right after the observation period in September 2022 and therefore allows for contextualization of these groups' further embeddedness in the German speaking Telegram network.

% \begin{figure*}[h]
% \centering
% \includegraphics[width=\textwidth]{images/messages_active_groups_peak.pdf}
% \caption{Number of messages (blue) and active authors (red) per day in the \textit{Schwurbelarchiv}. Maxima in messaging activity and related real-life events were identified manually.}
% \label{fig:content_w_peak}
% \end{figure*}

% \begin{figure*}[h]
% \centering
% \includegraphics[width=\textwidth]{images/messages_per_author_per_day.pdf}
% \caption{Number of messages in the time period between January 1, 2022 and the end of the observation period (blue) and number of messages posted per person per day (green).}
% \label{fig:content_per_person}
% \end{figure*}

\subsection{Transcription}
Uniquely, the dataset presented here contains transcriptions of both audio messages and video files. For the transcription process, we used the \textit{openai/whisper-large-v3-turbo} model, which was the latest \textit{whisper} model available at the time of publication~\cite{radford_robust_2023}. This increases the amount of text in the dataset from 1,761M words in the written messages to a total of approximately 2,718M words (see Table~\ref{tab:transcribe}), increasing the text corpus by 54.3\%. As media files comprise a considerable share of all forwarded messages (see also Tab.~\ref{tab:explo}), transcribed messages are an important carrier of information between chats. A feature of the text generated via transcription is that it consists of a well-defined vocabulary and therefore has no typos or alternative spellings. While this precludes the analysis of textual characteristics related to these features, it might allow for more accurate text analysis and topic identification and requires less pre-processing before further analysis. 
Regarding transcription quality, the Whisper large-v3-turbo model used for transcription achieves Word Error Rates (WER) between approximately 4\% and 7\% for German on standard benchmarks such as Common Voice 15 and Fleurs~\cite{radford_robust_2023}. However, these benchmarks consist of clean, read-aloud speech and therefore represent a best-case transcription accuracy. The audio content in the Schwurbelarchiv includes potentially incomprehensible speech features like informality, videos with background noise, and dialect speech, all of which are likely to increase error rates. To accommodate these challenges, we employed Voice Activity Detection (VAD) to filter non-speech segments, to help separate noise from non-voice signals. As for potential improvements: speaker diarization and the use of the full (non-turbo) Whisper model would have offered potential quality improvements but were out of scope due to the substantially higher computational cost. We consider the chosen pipeline to represent the best trade-off between transcription quality and feasibility at this scale. We did not conduct a manual evaluation of transcription quality on a sample of the data, which constitutes a limitation of this work.

\begin{table}[ht]

\centering

% \resizebox{\columnwidth}{!}{%
% \begin{tabular}{|c|cc|}
% \hline
%                 & Video & Voice Message \\ \hline
% Sum filesizes   &                            &    397.24\,GB \\ \hline
% Sum duration    &                            &    27,803 h   \\ \hline
% Median duration &                            &    32.48 sec  \\ \hline
% \end{tabular}
% }

% \resizebox{\columnwidth}{!}{%
\begin{tabular}{|c|cccc|}
\hline
                &  Text&Voice messages     & Video      & Images    \\ \hline
Count           &  54.10M&0.87M& 1.58M& 11.87M\\ \hline
Total words     &  1,761M&530M& 427M&           \\ \hline
Total filesize  &  &0.91 TB& 17.43 TB& 1.14 TB\\ \hline
Total duration  &  &65,433 h& 61,406 h&           \\ \hline
Median duration &  &76.83 s& 74.00 s&           \\ \hline
\end{tabular}
\vspace{3mm}

\caption{Multimodal data statistics. Excludes video and audio files that were corrupted, did not include an audio track or had an unknown codec.}
\label{tab:transcribe}
\end{table}

Next to voice messages and video files, 78\% of the 15.3M media files that were sent by Telegram users in the dataset were images. Manual investigation of a small fraction of the images suggests that many of them contain text in the form of memes, slogans, or announcements. While we believe it is technically feasible to extract text from these images or even use new multimodal large language models to create text descriptions of them, such processing incurs additional computational cost and is left to future work. We note that the images are not part of the dataset released by us within the scope of this work.

\section{Exploratory Analysis}

To illustrate the kind of analysis the \textit{Schwurbelarchiv} enables, we provide an analysis of overall posting activity and activity distribution across authors and chats.

\subsection{Overall Posting Activity}
\begin{figure}[ht]
\centering
\includegraphics[width=\columnwidth]{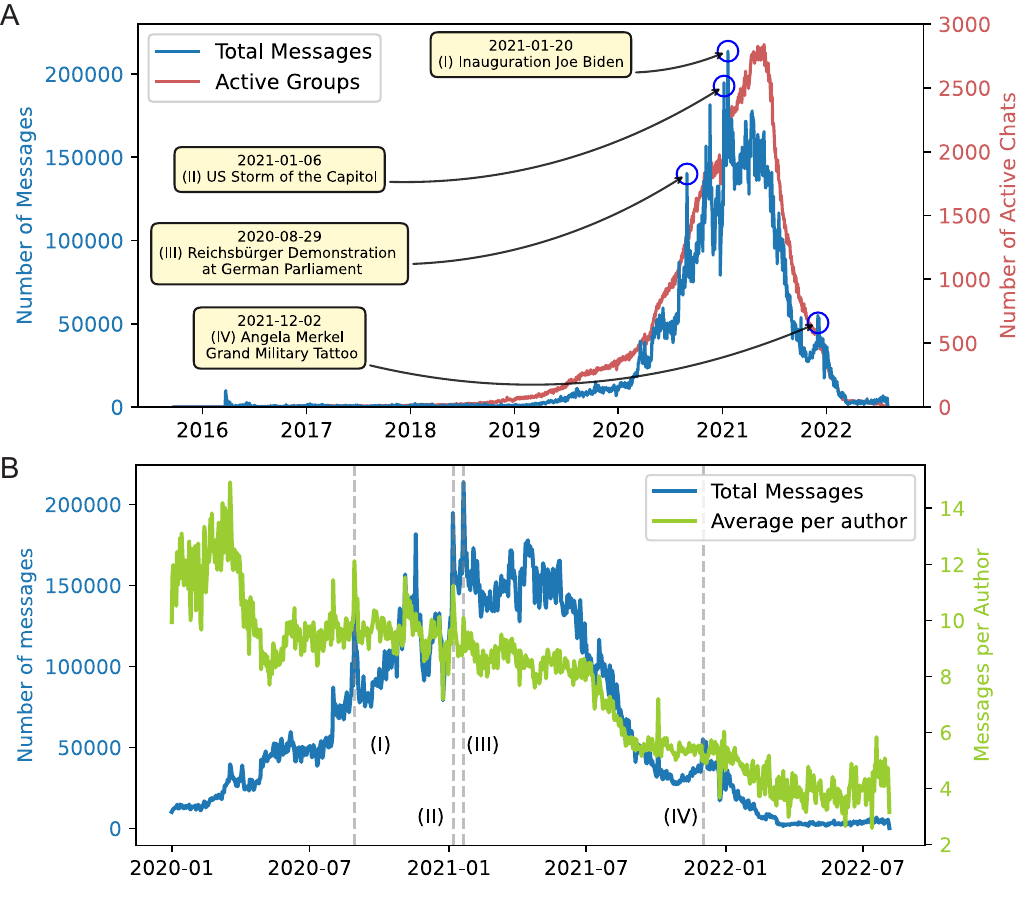}
\caption{
\textbf{A.} Number of messages (blue) and active chats (red) per day in the Schwurbelarchiv. Maxima in messaging activity and related real-life events were identified manually.
\textbf{B.} Number of messages in the time period between January 1, 2022 and the end of the observation period (blue) and number of messages posted per person per day (green). The numbers in parentheses indicate the position of the events highlighted in panel A.
}
\label{fig:timeseries}
\end{figure}

Figure~\ref{fig:timeseries}A shows the overall posting frequency in terms of messages and active chats over the full observation period. Active chats in this context are defined as chats that had at least one posted message on a given day. The number of active chats and message volume already started to increase slightly before the onset of the COVID-19 pandemic. However, the dramatic increase in active chats and messages coincides with the onset of the pandemic in Europe. Peaks in message volume coincide temporally with political events such as a large demonstration at the German parliament on August 29, 2020, and the Capitol Riot in the U.S. on January 6, 2021. However, these associations are based on temporal co-occurrence and have not been verified through content analyses. Further insight into the discussion topics at these times could be gained for example by topic modelling of the messages posted during these events. 

In Fig.~\ref{fig:timeseries}B we again show the overall number of messages per day, overlaid with the number of messages per author and day. After an initial period of high activity in early 2020 with around 13 messages per author and day we observe a steep decline of individual author activity in April and May 2020 that does not coincide with a decline in the overall posting traffic. This indicates that while individual authors became less active, more authors joined the discourse on Telegram, contributing to a steady increase of overall message volume. This is also consistent with our impressions from deep-reading small fractions of the content: we find that as conspiracy-related Telegram chats gain in popularity, established authors are replaced by newcomers that want to learn about conspiracy theories like \textit{QAnon}, \textit{Pizzagate}, or \textit{Chemtrails}, but rarely establish themselves as experts. We call this behaviour \textit{Conspiracy Tourism} which might be an interesting avenue for further, more qualitative research.
We observe a more gradual decline in author activity from June 2020 to June 2021 where authors post on average nine messages a day, followed by another steep decline in late summer 2021. At the end of our observation period, e.g. in summer 2022, the activity of authors seems to have stabilized around three messages per author and day. 

\subsection{Activity distribution}
A common aspect of social media platforms is high activity concentration, where few drivers (authors, chats) are responsible for most of the content/attention~\cite{orellana-rodriguez_attention_2018}. The \textit{Schwurbelarchiv} is no different. Grouping the messages by unique authors and chats (Fig.~\ref{fig:dataset_statistics}C), we see that the top 10\% of chats aggregate 73.3\% of all messages and the top 10\% authors contribute 94.6\% of all messages. This is reflected in the very high Gini indexes of \mbox{G=0.83} (chats) and \mbox{G=0.95} (authors). For comparison, Gini coefficients reported for attention metrics on Twitter are in a similar range with G=0.94 for followers and G=0.90 for retweets~\cite{zhu_attention_2016}. It is, however, important to mention that these measure different quantities than posting activity per chat and are therefore not directly comparable. For Telegram and other messaging platforms, comparable activity concentration metrics are, to our knowledge, not available in the published literature.

The probability distributions of messages and unique authors per chat (Fig.~\ref{fig:dataset_statistics}D) have very wide ranges, with the largest chats having on the order of $10^4$ authors and $10^6$ messages. Due to the uncertainty around the dataset sampling scheme, we refrain from e.g. fitting power-law distributions as the sampling can heavily bias both measured exponents and model selection of the best-fitted distribution (e.g. power-law vs. lognormal). Regardless, the distributions demonstrate fat or heavy tails.

\section{Discussion}
%Rather than relying on data collection from the Telegram API, the dataset was scraped with an automated approach via Telegram desktop clients that enabled the download of multimedia content, e.g. images, videos and voice messages. 
The \textit{Schwurbelarchiv} is a dataset of content from 5,807 predominantly German-language Telegram chats in which conspiracy-related topics. 
The majority of content contained in the dataset was produced during the first two years of the COVID-19 pandemic. 
Uniquely, next to text messages the dataset also captures a large part of the multimedia content shared in the observed chats. 
Such content constitutes 16.7\% of all original messages, demonstrating that multimedia content indeed plays an important role.
What is more is the outsized importance of such content for information transmission between chats, where 42.2\% of forwarded messages contain multimedia content.
Based on our qualitative observations of Telegram chats, the discussion is often driven by video or voice messages while text messages are mostly used to comment on this multimedia content. 
Therefore, by ignoring multimedia content -- as is often-times the case in research that uses digital trace data -- large parts of the discussion are potentially missed.
The -- to our knowledge unique -- contribution of the presented dataset is the provision of transcriptions of voice messages and videos alongside text messages and message metadata. 
Transcribing the more than 126,000 hours of audio messages and videos increases the number of words in the corpus available for text-based computational analysis by over 50\%. 
Our work therefore for the first time enables research into the role of multimedia content and the interplay between multimedia content and text messages in conspiracy-related discourse and adjacent fields such as misinformation and political extremism.

Next to the content of the messages, the data offers insights into the dynamics of information sharing and highlights how message frequency, group activity, and author engagement are distributed and evolve over time. The activity of authors in groups exhibits heavy-tailed distributions. This is consistent with activity patterns across social media platforms, such as Twitter~\cite{orellana-rodriguez_attention_2018}. Furthermore, activity distribution is very uneven, with the top 10\% chats aggregating 73.3\% of all messages and the top 10\% authors contributing 94.6\% of all messages.
Posting activity in the dataset peaks during the COVID-19 pandemic and at key political events of the time, indicating heightened and potentially coordinated online activity in the special interest groups that converse in the covered chats.
The metadata (time stamps, forwarding information, contained files) contained in the \textit{Schwurbelarchiv} allow for an empirical study of such activity.

Telegram does not provide a searchable index and therefore no clear picture of the sampled population is available.
This introduces severe limits to data collection: only manual curation of chats or snowball sampling are possible. 
We assess completeness of the sampled data in relation to all German-language conspiracy-related discourse happening on Telegram during the observation period by employing other means: the overlap of chats with similar datasets, and the fraction of forwarded messages whose origin is also contained in the dataset.
Snowball sampling, the approach employed to identify the chats contained in the \textit{Schwurbelarchiv}, is dependent on the set of initial chats.
Furthermore, volatility in user activity and the dynamic emergence of new chats complicate the sampling.
This is reflected in the diverging coverage of chats when comparing the \textit{Schwurbelarchiv} to other data sets collected in a similar way~\cite{baumgartner_pushshift_2020, golovin2026teragram, zehring_german_2023}, and when comparing these datasets among each other.
While the low overlap in coverage is not a limitation in itself, the substantial influence of temporal and cultural context and the initial chats from which the sampling is started need to be kept in mind for downstream analyses.
Looking at forwarded messages we find that 54.1\% of incoming forwarded messages are also contained as original messages in the dataset. On the flip side, this means that around half of the forwarded messages originate in chats that are not contained in the \textit{Schwurbelarchiv}. We therefore estimate that the \textit{Schwurbelarchiv} covers around half of the conspiracy-related discourse happening in predominantly German-language Telegram chats at the time. Other than potentially undersampling smaller chats from which fewer messages are forwarded, we have no reason to believe that the selection of chats included in the dataset is systematically biased. 
Furthermore, we have validated the high-level chat selection criteria employed by the original curator of the data set by applying a machine-learning based classifier to individual messages longer than 150 characters to classify them as conspiracy-related.
We find that for a substantial number of chats (18.4\%) the classifier detects no conspiracy content, these chats only account for 0.5\% of all messages longer than 150 characters.
We therefore believe that the \textit{Schwurbelarchiv} is an adequate resource for the study of conspiracy theories in the given temporal and geographic scope as it provides both good coverage of the discourse at the time as well as a targeted selection of relevant chats.

\subsection{Limitations}
In terms of information contained in the \textit{Schwurbelarchiv}, one substantial limitation of this dataset is the lack of unique user identifiers. As users can freely choose their usernames, the lack of access to user IDs makes it impossible to determine with certainty whether different authors are truly distinct individuals or whether posts with the same username originate from the same user. However, we believe the assumption that different usernames largely indicate activity by different individuals is warranted.

Another limitation of the \textit{Schwurbelarchiv} dataset relates to the unknown point in time at which data was collected for each chat. As has been reported by~\citet{buehling_message_2024}, messages on Telegram are deleted relatively frequently, and deletion patterns are not random. We estimate large portions of the \textit{Schwurbelarchiv} dataset, particularly towards the end of the observation period, to have been collected via streaming and therefore be relatively unaffected by message deletion. However, we cannot determine the point at which streaming started for each chat. Nevertheless, we note that the other datasets covering a similar geographical and temporal scope have all been collected by bulk download through the Telegram API~\cite{zehring_german_2023, golovin2026teragram, baumgartner_pushshift_2020}. Therefore, these datasets are subject to increasing message deletion the farther a point in time is away from the time of data collection as well. By including both bulk-downloaded and streamed portions of chat content, the \textit{Schwurbelarchiv} is likely more complete in terms of deleted messages than other datasets.

A further, conceptual limitation concerns the construct of the conspiracy theory itself: no universally agreed-upon definition exists, even among researchers~\cite{uscinski2023conspiracy, douglas2023conspiracy}, and computational detection methods diverge accordingly~\cite{diab2024classifying}. We adopt the working definition of~\citet{uscinski2018}, operationalised through the in-domain classifier of~\citet{pustet_detection_2024}, as a serviceable but not definitive choice; as the field and its definitions keep evolving, the \texttt{is\_conspiracy} label in the dataset should be read as one in-domain operationalisation rather than as ground truth, whose reliability for conspiracy narratives beyond the cultural and temporal context is not guaranteed~\cite{fort2023bigfoot}. We further note that the plausibility of a conspiracy theory does not enter this definition. We do not believe it is feasible to assess the plausibility of a conspiracy theory in an automated way and in temporal proximity to the events it addresses. The \texttt{is\_conspiracy} label therefore does not distinguish between genuine conspiracies and conspiracy theories.

\subsection{Future Work}
To address the lack of unique user identifiers, we plan to investigate text features and variance in temporal messaging patterns in the future to distinguish between authors who may share the same name, addressing one of the primary limitations of the \textit{Schwurbelarchiv}. 
Furthermore, while the raw dataset contains a large number of images, these are currently not accessible to text-based research approaches.
We plan to employ image-to-text models to create suitable descriptions of the images that can be analysed with such workflows.
Next to the content of messages, the \textit{Schwurbelarchiv} data can also be analysed from a network perspective by treating groups or authors as nodes and linking them according to shared membership or participation. This approach reveals topological features such as common affiliations or frequently co-occurring authors. Future research could harness these networks to investigate community structures, track the spread of (mis-)information and narratives, and identify influential participants across different discussions. More broadly, future work can build on the dataset to investigate the dynamics of misinformation spread, conduct topic modelling, explore the influence of real-world events on user behaviour in online communities, and analyse the dynamics of conspiracy-related discourse in individual chats.

\section{Conclusion}
In this paper we present the \textit{Schwurbelarchiv} dataset, a large-scale collection of more than 5,800 Telegram chats, focusing primarily on German-language conspiracy-related discussions in the time from 2015 to 2022. With the present data publication we provide a cleaned and validated version of the dataset including transcribed audio content that is easy to access and re-use for other researchers. We clarify questions around the ethics of using the dataset as well as the terms of use. The dataset contains digital trace data of individuals. While we pseudonymized usernames, the dataset is not completely anonymised and in principle allows for the re-identification of individuals based on publicly disclosed information contained in the data. We release this dataset because we believe the benefit of being able to study the spread of conspiracy theories in a completely unregulated online environment at scale during a time of crisis outweighs the risk of infringing on user expectations and privacy. In particular, what we add with this publication is data cleaning and validation of a dataset that was originally collected by an anonymous data hoarder. Malicious use that attempts to single out individuals is already possible with the original dataset available on the Internet Archive. We emphasize that we do not intend this dataset to be used for the profiling of individuals and strongly caution against any attempts of re-identification.
However, when utilized for aggregate analyses, the \textit{Schwurbelarchiv} respects user expectations and provides broad utility for research in social network analysis, linguistic analysis, political science, and information studies, particularly in the realm of conspiracy theories.

\backmatter

%\bmhead{Supplementary information}

%\bmhead{Acknowledgements}
%\input{content/acknowledgements}
% !TEX root =  ../main.tex

\section*{Declarations}
\bmhead{Availability of data and materials}
The original \textit{Schwurbelarchiv} dataset is available at the Internet Archive\footnote{https://archive.org/details/schwurbel-archiv}. The metadata of the cleaned and processed dataset described in this paper -- that is, all data columns except the four text fields: the content of messages and of forwarded messages, and the transcriptions of speech-containing media in messages and in
forwarded messages -- is openly available on Zenodo\footnote{https://doi.org/10.5281/zenodo.21162732} and is sufficient to reproduce all results reported here. To protect the privacy of the individuals whose communications the dataset contains, the full version, which additionally includes these four text fields, is published as a separate, access-restricted record on Zenodo\footnote{https://doi.org/10.5281/zenodo.21161579}; access for research purposes can be requested directly through the Zenodo record, and the authors grant it on a case-by-case basis. All code used to parse and analyse the data is publicly available on GitHub\footnote{https://github.com/cs2-lab/schwurbelarchiv}.
\bmhead{Competing interests}
The author declare no competing interests.

\bmhead{Funding}
This research was funded in whole or in part by the Austrian Science Fund (FWF, DOI:10.55776/P37280). For open access purposes, the author has applied a CC BY public copyright license to any author-accepted manuscript version arising from this submission. MA, EH \& JP acknowledge the FWF funding.

\bmhead{Authors' contributions}
MA, JL \& JP conceived and designed the study. MA, EH, JL \& JP performed data analysis and cleaning. MA, EH, JL \& JP drafted the manuscript. All authors read and approved the final manuscript.

\bmhead{Acknowledgments}
The authors thank the anonymous individual who collected the original dataset, which made this work possible.

\newpage
 \begin{appendices}

% \section{Section title of first appendix}\label{secA1}
% !TEX root =  ../main.tex

\section{Column description of the dataset}

\begin{table}[htbp]
\centering
\begin{tabular}{l | p{10cm}}
\textbf{Column} & \textbf{Description} \\ \hline
uuid & unique identifier for each row \\ \hline
folder\_id & internal folder identifier for the group/channel \\ \hline
chat\_type & type of chat (\texttt{channel} or \texttt{group}) based on number of different authors within a group ($>3$ authors == group); includes manual classification for 37 chat\_names \\ \hline
chat\_name & name of the group/channel \\ \hline
author & hashed name of author \\ \hline
link\_url & link as found in HTML \texttt{a}-tag \\ \hline
media\_file & name of attached media file \\ \hline
media\_file\_type & \texttt{voice message}, \texttt{video}, or \texttt{photo} \\ \hline
message & message content \\ \hline
message\_id & ID of message in respective group \\ \hline
posting\_date & date and time of message in the group (UTC-aware \texttt{timestamptz}) \\ \hline
replied\_to & message\_id of message that has been replied to \\ \hline
website & website domain, if link\_url refers to a website \\ \hline
fwd\_link\_url & link from forwarded message \\ \hline
fwd\_media\_file & internal path of the forwarded media file \\ \hline
fwd\_media\_file\_type & \texttt{voice message}, \texttt{video}, or \texttt{photo} \\ \hline
fwd\_message & forwarded message content \\ \hline
fwd\_posting\_date & date and time of message in the original group \\ \hline
message\_lang & language detected in message \\ \hline
is\_conspiracy & whether message was classified as conspiracy content \\ \hline
audio\_transcription & transcribed content of a speech-containing audio file \\ \hline
audio\_length\_s & duration of an audio file (in seconds) \\ \hline
audio\_size\_mb & file size of the attached audio file (in MB) \\ \hline
audio\_lang & language detected in audio transcription \\ \hline
video\_transcription & transcribed content of a speech-containing video file \\ \hline
video\_length\_s & duration of a video file (in seconds) \\ \hline
video\_lang & language detected in video transcription \\
fwd\_posting\_date & date and time of message in the original group (UTC-aware \texttt{timestamptz}) \\ \hline
fwd\_author & hashed name of the original author of the forwarded message \\ \hline
is\_duplicate & boolean flag marking rows that duplicate another due to redundant scraping of that group. Lines kept and flagged instead of deleted as for some semantic duplicates, the replied\_to column references to only one of these lines.\\ \hline
\end{tabular}
\caption{Schwurbelarchiv dataset columns.}
\label{tab:columns}
\end{table}

\newpage

\section{Base list of English keywords by \cite{mohr_inference_2023}}

\label{app:keywords}
\texttt{keywords = \{ 'covid', 'corona', 'virus', 'pandemic', 'lockdown',
'health', 'mask', 'distancing', 'outbreak', 'symptom',
'quarantine', 'influenza', 'vaccine', 'vaccination', 'pandemic',
'ventilator', 'isolation', 'immunity', 'hospital', 'icu',
'intensive care unit', 'treatment', 'virologist', 'clinic',
'homeopathy', 'sick', 'pharma', 'polymerase chain reaction',
'pcr', 'pertussis', 'emergency', 'injection', 'doctors',
'cellular', 'remote work', 'frontline', 'covid-19', 'sars',
'sanitation', 'pathogen', 'propaganda', 'disease', 'epidemic',
'diarrhea', 'adjuvants', 'respiratory', 'hygiene', 'protein',
'medicine', 'new cases', 'positive', 'scientist', 'contagious',
'mandate', 'variant', 'infect', 'deaths', 'ill', 'cough',
'measure', 'viral', 'mrna', 'prevent', 'healthcare', 'contract',
'shutdown', 'smallpox', 'booster', 'antibody', 'dose', 'evidence',
'misinformation', 'isolation', 'observed', 'mandatory',
'allergic', 'allergy', 'immune', 'shortage', 'syndrome', 'drug',
'chinese', 'test', 'restriction', 'spread', 'vax', 'experimental' \}}

\newpage

% %%=============================================%%
% %% For submissions to Nature Portfolio Journals %%
% %% please use the heading ``Extended Data''.   %%
% %%=============================================%%

% %%=============================================================%%
% %% Sample for another appendix section			       %%
% %%=============================================================%%

% %% \section{Example of another appendix section}\label{secA2}%
% %% Appendices may be used for helpful, supporting or essential material that would otherwise 
% %% clutter, break up or be distracting to the text. Appendices can consist of sections, figures, 
% %% tables and equations etc.

\end{appendices}


\bibliography{content/references_arxiv}% common bib file

\end{document}